\documentclass[preprint,numafflabel]{elsarticle}
\usepackage[top=1.7cm, bottom=2cm, outer=2.5cm, inner=2.5cm]{geometry}
\usepackage[english]{babel}
\usepackage{lineno,hyperref}
\usepackage{multicol}

\usepackage{mathtools}
\usepackage{amsmath}
\usepackage{amssymb}
\usepackage{widetext} 

\usepackage{graphicx}
\usepackage{float}
\usepackage{caption}

\usepackage{soul}
\usepackage{xcolor}
\newcommand{\mini}{\scriptscriptstyle}

\journal{some nice journal}
\modulolinenumbers[5]

\begin{document}


\begin{frontmatter}

\title{Comments on the anti-Hawking effect on a BTZ black hole spacetime}
\author[1,2]{Lissa de Souza Campos}
\ead{lissa.desouzacampos01@universitadipavia.it}
\author[1,2]{Claudio Dappiaggi}
\ead{claudio.dappiaggi@unipv.it}
\address[1]{Dipartimento di Fisica, Universit\`a degli Studi di Pavia, Via Bassi, 6, 27100 Pavia, Italy}
\address[2]{Istituto Nazionale di Fisica Nucleare -- Sezione di Pavia, Via Bassi 6, 27100 Pavia, Italy}

\begin{abstract}
	We compute the transition rate of an Unruh-DeWitt detector coupled both to a ground state and to a KMS state of a massless, conformally coupled scalar field on a static BTZ black hole with Robin boundary conditions. We observe that, although the anti-Hawking effect is manifest for the ground state, this is not the case for the KMS state. In addition, we show that our analysis applies with minor modifications also to the anti-Unruh effect on Rindler-AdS$_3$ spacetime.
\end{abstract}


\end{frontmatter}

\begin{multicols}{2}


\section{Introduction}


	Particle detectors are a valuable theoretical tool to probe the state of a quantum field. Particularly noteworthy is the one known as Unruh-DeWitt detector, which is modeled as an idealized atom with a two-level energy system interacting with a free real scalar field by means of a monopole coupling \cite{Unruh,DeWitt}. Thanks to its versatility, in the past years it has been used in several contexts, especially in connection to the analysis of the Unruh effect and of Hawking radiation, see {\it e.g.} \cite{Hodgkinson:2012mr,Louko:2007mu,Louko:2006zv,Ng:2014kha}. As explained in detail in these references, the fundamental quantity of interest in this model is the probability of a transition between the energy eigenstates, which is computed at first order in perturbation theory.\\
	\indent Recently, it has been shown that Unruh-DeWitt detectors can measure apparently counter-intuitive effects dubbed ``anti-Unruh effect'' and  ``anti-\-Haw\-king effect'' \cite{Henderson:2019uqo,Brenna:2015fga,garay2016thermalization}. The standard reasoning is that a non-inertial observer perceives a vacuum state as a thermal bath with temperature proportional to its acceleration. One could expect that this implies that also the response of a detector interacting with this state is proportional to the detector proper acceleration. In general, this is not the case. When a detector ``clicks less at positions of higher temperature'', we call it anti-\-Unruh or anti-\-Haw\-king effect, depending on the specific underlying scenario. The existence of these phenomena is deduced by analyzing the transition rate of the Unruh-DeWitt detector, which, in turn, depends critically on the two-point correlation function of the underlying matter field. \\
  %
	\indent	In this paper we study the emergence of the above-mentioned phenomena by considering an Unruh-DeWitt detector traveling on static trajectories on a BTZ black hole, and on its universal cover, Rindler-AdS$_3$, interacting with both a ground state and a KMS state of a real, massless, conformally coupled scalar field. These correspond, respectively, to the Boulware and Har\-tle-\-Haw\-king-\-Is\-rael states in the exterior region ($r>r_h$).
	Yet, since the underlying spacetimes are asymptotically AdS, contrary to globally hyperbolic spacetimes, the construction of a ground as well as that of a KMS state at fixed temperature is not unique. One needs to specify boundary conditions for the underlying field which affect drastically the form of the two-point function. In particular, we are interested in considering boundary conditions of Robin type and we discuss them following a procedure which has been thoroughly investigated in \cite{Dappiaggi:2018xvw,Dappiaggi:2018pju,Dappiaggi:2016fwc,Bussola:2018iqj,CD20}---see also \cite{Garbarz:2017wzv,Ishibashi:2004wx}, for related analyses.\\
	\indent Our results expand previous works on BTZ spacetime by considering and comparing the two different states and by including more general boundary conditions \cite{Hodgkinson:2012mr,Henderson:2019uqo}.
	Hence, on the one hand, we are able to extend the results of \cite{Henderson:2019uqo,Brenna:2015fga} to all boundary conditions of Robin type. On the other hand, we observe that neither the anti-Unruh nor the anti-Hawking effect is manifest for the KMS states.\\
	\indent First, in Section \ref{sec:Geometry}, we outline the geometry of the spacetimes considered. In Section \ref{sec:transition}, we describe how to obtain the two-point functions and write an explicit expression for the transition rate. The numerical analyses and main results are in Section \ref{sec:numerical}. 
	%
 	%

\section{Geometric data}
\label{sec:Geometry}

\begin{figure}[H]
		\centering
				\includegraphics[width=.25 \textwidth]{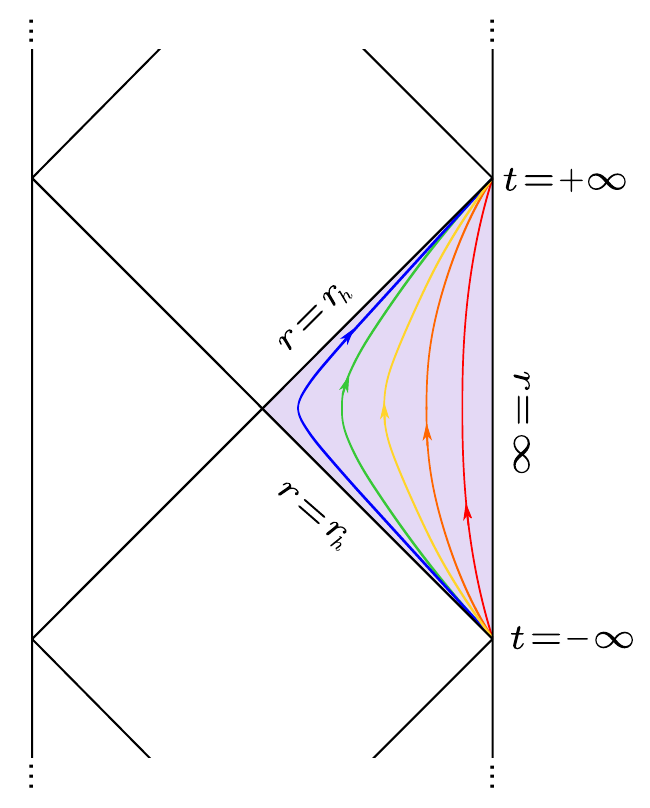}
		\caption{Rindler-AdS$_3$ spacetime and its static trajectories highlighted in the corresponding region on the conformal diagram of AdS$_3$.}
		\label{fig:frog2HAHAHA}
		\end{figure}

	In this work we consider two distinguished solutions of vacuum Einstein's equations with a negative cosmological constant $\Lambda$ normalized to $-1$, namely a static BTZ black hole \cite{Banados:1992wn} and its universal cover, Rindler-AdS$_3$ spacetime (or wedge) \cite{Parikh:2012kg}. Their associated line element reads
		\begin{equation}
			\label{eq:metric}
				ds^2=-(r^2-r_h^2)dt^2+(r^2-r_h^2)^{-1}dr^2+r^2d\theta^2,
		\end{equation}
	where $t\in\mathbb{R}$ while $r_h>0$ is a constant which, in the case of BTZ spacetime, encodes the information of the black hole mass $M=r^2_h$. In addition, $r\in (r_h,\infty)$ and $\theta$ runs over the whole real line on the Rindler-AdS$_3$ background, while $\theta\in[0,2\pi)$ in the case of a BTZ black hole. Focusing on the Rindler-AdS$_3$ wedge, notice that the surface $r=r_h$ is an observer dependent acceleration horizon seen by supercritically accelerated observers with proper acceleration $a = r(r^2-r_h^2)^{-1/2}>1$. Their trajectories correspond to complete integral lines of the Killing field $\partial_t$, see Figure \ref{fig:frog2HAHAHA}.
	The Killing vector $\xi=\partial_t$ generates a bifurcate Killing horizon at $r=r_h$ and accordingly, we define the local Hawking temperature as
		\begin{equation}
			\label{eq:temperature}
				T_H := \frac{\kappa_h}{2\pi \sqrt{|g_{00}|}},
		\end{equation}
	where $\kappa_h=r_h$ is the surface gravity computed at the horizon. The Hawking temperature as per Equation \eqref{eq:temperature} is defined so to take into account time-dilations effects introduced by the $r$-dependence of $g_{00}$. As a consequence, it diverges at the horizon and it vanishes at infinity. This is in sharp contrast with the counterpart in asymptotically flat black hole spacetimes which tends to a finite quantity at infinity. At the same time, it does coincide with the effective Unruh temperature obtained by the global embedding on a higher-dimensional flat spacetime procedure, that is: $T_H=\frac{1}{2\pi}\sqrt{a^2 - 1}$, with $a$ defined as before \cite{deser1997accelerated,deser1999mapping}.


\section{The transition rate}
 \label{sec:transition}


In this section we consider an Unruh-DeWitt detector either on Rindler-AdS$_3$ or on a static BTZ spacetime following static trajectories pa\-ram\-e\-triz\-ed by their proper time $\tau$ and at fixed spatial coordinates $\underline{x}$. Observe that, from Equation \eqref{eq:metric}, one can infer $d\tau=\sqrt{r^2-r^2_h}dt$. The detector interacts with a real, massless, conformally coupled scalar field. Following the standard theory, see {\it e.g.} \cite{Hodgkinson:2012mr,Louko:2006zv,Fewster:2016ewy}, in the infinite interaction time limit, the instantaneous transition rate can be evaluated as the Fourier transform of the pull-back along the detector trajectory of the two-point function $\omega_2(\tau,\underline{x},\tau^\prime,\underline{x}^\prime)$ of the underlying field. In other words, recalling that the underlying spacetime is static and that the detector lies at fixed spatial coordinate, $\underline{x}=\underline{x}^\prime$, the transition rate reads:
\begin{align}
	\label{eq:transitionjustlike33}
	\dot{\mathcal{F}} &= \int_\mathbb{R}ds e^{-i\Omega s} \omega_2(s),
\end{align}
where $s=\tau-\tau^\prime$ while $\omega_2(s)$ is the above-mentioned pull-back of the two-point function of the underlying field. In the following, we find an explicit form for $\omega_2$ both for a ground and for a KMS state. Although we are interested both in Rindler-AdS$_3$ and BTZ spacetimes, we shall mainly focus on the latter. Since the procedure is almost identical in both scenarios, we focus on one highlighting only the necessary differences to translate our results to the other case.

Let $\Psi:M\to\mathbb{R}$ be a real scalar field where $M$ is either Rindler-AdS$_3$ or a static BTZ spacetime. The dynamics is ruled by
		\begin{equation}\label{Eq:KG}
			\left(\Box- \frac{3}{4}\right)\Psi=0,
		\end{equation}
where $\Box$ is the D'Alembert wave operator built out of Equation \eqref{eq:metric}. We consider solutions to (\ref{Eq:KG}) that can be written as linear combinations of separable solutions:
		\begin{equation}
			\label{eq:ansatz}
				\Psi(t,r,\theta)=\int_{\sigma(\Delta_1)}d\mu(\ell)\int_{\mathbb{R}}e^{-i\omega t} e^{i\ell\theta}R_{\omega,\ell}(r) ,
		\end{equation}
where $\Delta_1\equiv-\frac{\partial^2}{\partial{\theta^2}}$ is the one-dimensional Laplacian whose eigenfunctions are $e^{i\ell\theta}$. If the underlying spacetime is Rindler-AdS$_3$, $\ell$ runs over the real line and $\int_{\sigma(\Delta_1)}d\mu(\ell)=\int_{\mathbb{R}}d\ell$, while on BTZ, $\ell$ is integer valued and $\int_{\sigma(\Delta_1)}d\mu(\ell)=\sum\limits_{\ell\in\mathbb{Z}}$. Inserting Equation \eqref{eq:ansatz} in Equation \eqref{Eq:KG} it turns out that the remaining unknown is a solution of
		\begin{align}
			\label{eq:radialeq}
		 	\Bigg\{(r^2-r_h^2)&\partial^2_r + \left( 3r - \frac{r_h^2}{r} \right) \partial_r \nonumber\\
				& - \frac{\ell^2}{r^2} +\frac{\omega^2}{(r^2-r_h^2)} - \frac{3}{4} \Bigg\}R_{\omega,\ell}(r) = 0,
		\end{align}
In order to solve Equation \eqref{eq:radialeq}, it is convenient to introduce the coordinate $z= \frac{r^2-r_h^2}{r^2}\in(0,1)$ and, using Frobenius method to infer the behaviour of $R_{\omega,\ell}(r)$ close to the endpoints $z=0,1$, we set
		\begin{equation}
			\label{eq:ansataztohyper}
			R_{\omega,\ell}(z) =z^\alpha (1-z)^\frac{3}{4} v_{\omega,\ell}(z),
		\end{equation}
	where $\alpha=i\frac{\omega}{2r_h}$. The function $v_{\omega,\ell}(z)$ is in turn a solution of the hypergeometric differential equation
	\begin{equation}\label{eq:hypergeom}
	z(1-z)\frac{d^2 v}{dz^2}+(c-(a+b-1)z))\frac{dv}{dz}-ab v=0,
	\end{equation}
	where we have dropped, for simplicity of notation, the subscripts $\omega,\ell$ and where, setting $\Upsilon_\ell\doteq i\frac{\ell}{ 2r_h}$,
	\begin{subequations}
		\label{eq:Coefabc}
		\begin{align}
			a &= \alpha +\frac{3}{4} + \Upsilon_\ell	, \label{eq:a}\\
			b &=  \alpha + \frac{3}{4} -	\Upsilon_\ell, \label{eq:b}\\
			c &= 1 + 2\alpha. \label{eq:c}
		\end{align}
	\end{subequations}
A reader who is comparing the above construction with the existing literature should keep in mind that we are adopting slightly different conventions in comparison with \cite{BDFK}. As discussed thoroughly in \cite{BDFK}, a criterion to select a specific solution of Equation \eqref{eq:hypergeom} consists of assigning a boundary condition of Robin type at $z=1$, following the general theory of Sturm-Liouville ordinary differential equations, {\it cf.} \cite{Zettl:2005}.\\In other words, we consider
	\begin{subequations}
		\label{eq:solutions}
		\begin{align}
				&R_{\omega,\ell,\gamma}(z)=z^\alpha (1-z)^{\frac{3}{4}} \big(\cos(\gamma) v_{1(1)}(z)  \nonumber \\
				&\hspace{4cm} + \sin(\gamma) v_{2(1)}(z)\big), \label{eq:robin} \\
				&\text{where} \nonumber \\
				&v_{1(1)}=F(a,b;a+b+1-c;1-z),\label{eq:sol11hypergeo}\\
				&v_{2(1)}=(1-z)^{c-a-b} \nonumber \\
				&\qquad \quad\times F(c-a,c-b;c-a-b+1;1-z).\label{eq:sol21hypergeo}
			\end{align}
		\end{subequations}
	In Equation \eqref{eq:robin}, $\gamma$ is a parameter which codifies the Robin boundary condition assigned at the endpoint $z=1$ and here, we allow $\gamma\in[0,\gamma_c)$ where $\gamma_c= \arctan\left(-\frac{1}{2}\frac{\Gamma\left(1/4 + \Upsilon_\ell\right)\Gamma\left(1/4 - \Upsilon_\ell\right)}{\Gamma(3/4+ \Upsilon_\ell)\Gamma(3/4- \Upsilon_\ell)}\right)\in\left(\frac{\pi}{2},\pi\right).$ Observe that one discards values of $\gamma$ greater than $\gamma_c$ since they entail the existence of bound state modes in the two-point function, a scenario whose physical relevance is still under investigation. In addition, notice that $\gamma=0$ and $\gamma=\frac{\pi}{2}$ correspond respectively to Dirichlet and Neumann boundary conditions.

	As studied in detail in \cite{Bussola:2018iqj,BDFK} for the case of a BTZ spacetime, considerations from spectral theory allow us to start from Equation \eqref{eq:robin} to construct both the two-point function for the ground state and for a KMS state at inverse-temperature $\beta = 2\pi/r_h$. The same procedure used in these papers applies slavishly to the case of Rindler-AdS$_3$ spacetime changing accordingly the behaviour of the angular component of the two-point function. Hence we can limit ourselves to report the final result, namely
		\begin{widetext}
			\begin{equation}
				\label{eq:ansatzTWO}
				\omega_{2,0}(x,x')=\lim_{\epsilon\rightarrow 0^+}\int_{\sigma(\Delta_1)}\frac{ d\mu_\ell}{2\pi}\int_{\mathbb{R}} d\omega\,\Theta(\omega)e^{-i\omega (t-t'-i\epsilon)} \mathcal{N}_\ell R_{\omega,\ell,\gamma}(z)R_{\omega,\ell,\gamma}(z')     e^{i\ell(\theta-\theta^\prime)},
			\end{equation}

			\begin{equation}
				\label{eq:ansatzTWOKMS}
				\omega_{2,\mini{T_H}}(x,x')=\lim_{\epsilon\rightarrow 0^+} \int_{\sigma(\Delta_1)}\frac{ d\mu_\ell}{2\pi}\int_{\mathbb{R}} d\omega\Theta(\omega) \left[\frac{ e^{\omega/T_H}e^{-i \omega (t-t'-i\epsilon)}+e^{+i \omega (t-t'+i\epsilon)}}{e^{\omega/T_H}-1}\right]\mathcal{N}_\ell R_{\omega,\ell,\gamma}(z)R_{\omega,\ell,\gamma}(z')     e^{i\ell(\theta-\theta^\prime)}.
			\end{equation}
		\end{widetext}
		\noindent
Here $\Theta$ is the Heaviside function, while $\mathcal{N}_\ell$ is a normalization constant:
		\begin{align}
			\label{eq:normalization}
				\mathcal{N}_\ell= \frac{1}{i \pi}\frac{(\overline{A}B-A\overline{B})}{ |A\sin(\gamma) -B\cos(\gamma)|^2},
		\end{align}
	where the coefficients $A$ and $B$ are given by:
		\begin{subequations}
			\label{eq:CoefAeB}
				\begin{align}
					A&=\frac{\Gamma(c)\Gamma(c-a-b)}{\Gamma(c-a)\Gamma(c-b)}, \label{eq:CoefA}\\
					B&=\frac{\Gamma(c)\Gamma(a+b-c)}{\Gamma(a)\Gamma(b)}. \label{eq:CoefB}
				\end{align}
		\end{subequations}

Starting from Equations (\ref{eq:ansatzTWO}) and (\ref{eq:ansatzTWOKMS}), we obtain the transition rate as per Equation (\ref{eq:transitionjustlike33}) for an Unruh-DeWitt detector with energy gap $\Omega$ and following a static trajectory $x(\tau)=(\tau,z_{\mini{D}},\theta_{\mini{D}})$ where the subscript $D$ indicates that $(z_D,\theta_D)$ are the fixed spatial coordinates of the detector. With reference to the ground state, we obtain, for $\Omega<0$:
		\begin{align}
			\label{eq:transition}
				\dot{\mathcal{F}}_0=\int_{\sigma(\Delta_1)}\frac{ d\mu_\ell}{2\pi}\, \mathcal{N}_\ell R_{\omega,\ell,\gamma}(z_{\mini{D}})^2\Big|_{\omega=-\sqrt{|g_{00}|}\Omega},
		\end{align}
	while, for the KMS state at inverse-temperature $\beta = 2\pi/r_h$, we end up with
		\begin{align}
			\label{eq:transitionRindlerAdSKMS}
				\dot{\mathcal{F}}_{T_H}= \frac{\text{sign}(\Omega)}{e^{\text{sign}(\Omega)\omega/T_H}-1}\Big|_{\omega=+\sqrt{|g_{00}|}\left|\Omega\right|} \dot{\mathcal{F}}_0.
		\end{align}


\section{Numerical Analysis}
 \label{sec:numerical}


		In this section we state the main results of this paper based on the numerical analyses performed on the transition rate for both the ground state and the KMS state on both Rindler-AdS$_3$ and BTZ spacetimes. First, we study the $\ell=0$ contribution to the transition rate as a function of Hawking temperature, since $z_{\mini{D}}(T_{\mini{H}})=(1+4\pi^2T_{\mini{H}}^2)^{-1}$. Secondly, since the transition rate is given by an infinite sum in $\ell$, we analyze the behavior with respect to the truncation order, dubbed $\ell_{max}$.

		Notice that all plots are normalized with respect to its own maximum value. The numerical analyses summarized here is available at \cite{git_unruh_deWitt_BTZ}.


	\subsection{The detector response as a function of Hawking temperature}


			\begin{figure}[H]
			\centering
					\includegraphics[width=.45\textwidth]{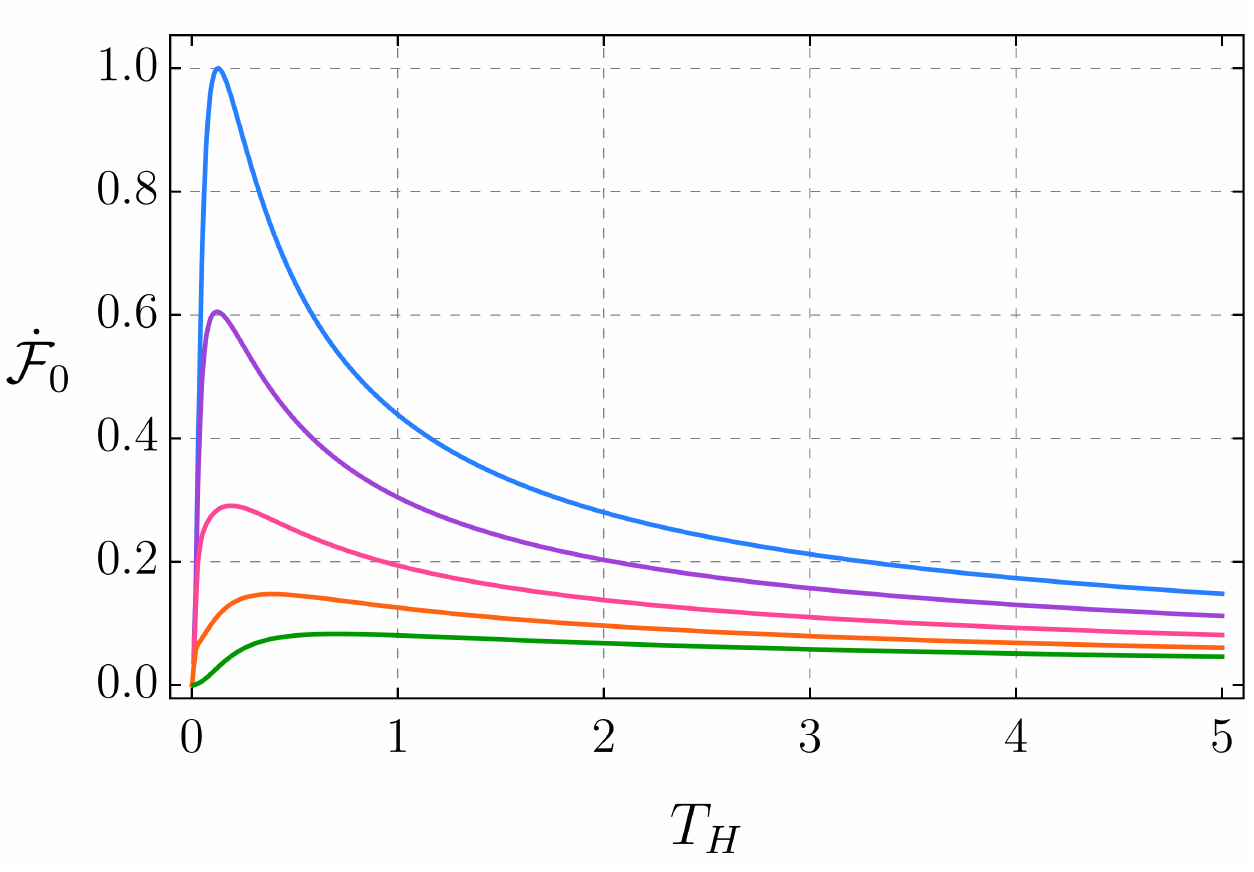}\hspace{.5cm}
			\caption{\label{fig:groundandkms1}$\ell=0$ contribution to the transition rate for the ground state as a function of the Hawking temperature for $r_h=1$, $\Omega=-0.1$ and different boundary conditions; from top to bottom, respectively, $\gamma=(0.50,0.47,0.40,0.25,0)\pi$. }
			\end{figure}
			\begin{figure}[H]
			\centering
					\includegraphics[width=.45\textwidth]{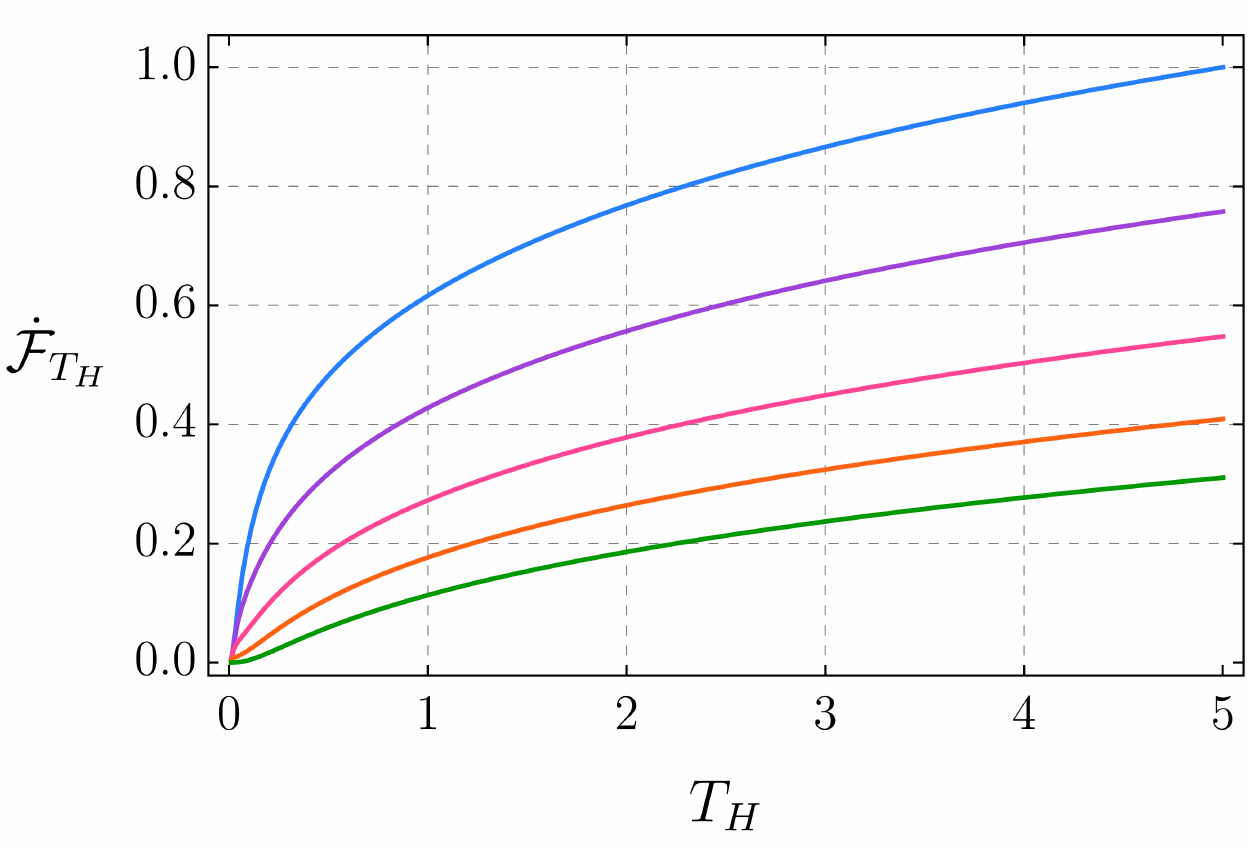}
			\caption{\label{fig:groundandkms2}$\ell=0$ contribution to the transition rate for the KMS state as a function of the Hawking temperature for $r_h=1$, $\Omega=-0.1$ and different boundary conditions; from top to bottom, respectively, $\gamma=(0.50,0.47,0.40,0.25,0)\pi$.}
			\end{figure}
		Since the only difference between Rindler-AdS$_3$ and the static BTZ case at the level of two-point function lies in the angular coordinate, the $\ell=0$ term of the detector response with respect to the Hawking temperature is the same in both cases. In other words, the outcomes presented in Figures \ref{fig:groundandkms1} and \ref{fig:groundandkms2} are valid for both Rindler-AdS$_3$ and for a static BTZ spacetime.

		Approaching the AdS$_3$ boundary, namely as
		$$r\rightarrow\infty \equiv z\rightarrow1 \equiv T_H\rightarrow0,$$
		for both states, the transition rate goes to zero for all boundary conditions. The detector response when coupled to the KMS state is compatible with the behavior of the Hawking temperature. As one can infer by direct inspection, both are increasing towards the horizon and diverging there, as $T_H\rightarrow\infty$, which is not the case for the ground state.

		Figures \ref{fig:groundandkms1} and \ref{fig:groundandkms2} show an interesting unexpected behavior regarding the contrast between the response of the detector when coupled to the ground state and when coupled to the KMS state. The anti-Unruh effect or anti-Hawking effect occurs when the derivative of the transition rate with respect to $T_H$ assumes negative values. It is worth emphasizing that $T_H$ is not the temperature of the Boulware-like ground state, but rather the temperature of the KMS state at the corresponding locus. As it emerges from Figure \ref{fig:groundandkms1}, we observe that, when coupled to the ground state, for the $\ell=0$ mode, the anti-Unruh effect is manifest for all boundary conditions. On the other hand, when interacting with the KMS state, as in Figure \ref{fig:groundandkms2}, for the $\ell=0$ mode, the detector responds with a monotonically increasing transition rate towards the horizon.

		In Figure \ref{fig:groundandkms1}, we notice that the effect is distinctly manifest for Neumann boundary condition, while it is almost imperceptible for Dirichlet boundary condition. For this reason, in the next section we include the analysis of the $\ell$-sum focusing on these two cases. However, the behaviour with respect to the $\ell$-sum does not affect much the behavior of the transition rate for the KMS states, in the sense that it continues to be a monotonically increrasing function with respect to $T_H$.

		We can draw the following conclusion. While, for the ground state, the results, already known in the literature, do extend to Robin boundary conditions, neither the anti-Unruh effect on Rindler-AdS$_3$ nor the anti-Hawking effect on a static BTZ black hole is observed by a detector interacting with the KMS state. Since KMS states are natural candidates to be used when describing thermal effects, this result encourages future investigations regarding the relation between the effects and KMS states.


	\subsection{With respect to the $\ell$-sum}


		\paragraph{The scenario $r_h=1$}

\begin{figure}[H]
				\centering
						\includegraphics[width=0.45\textwidth]{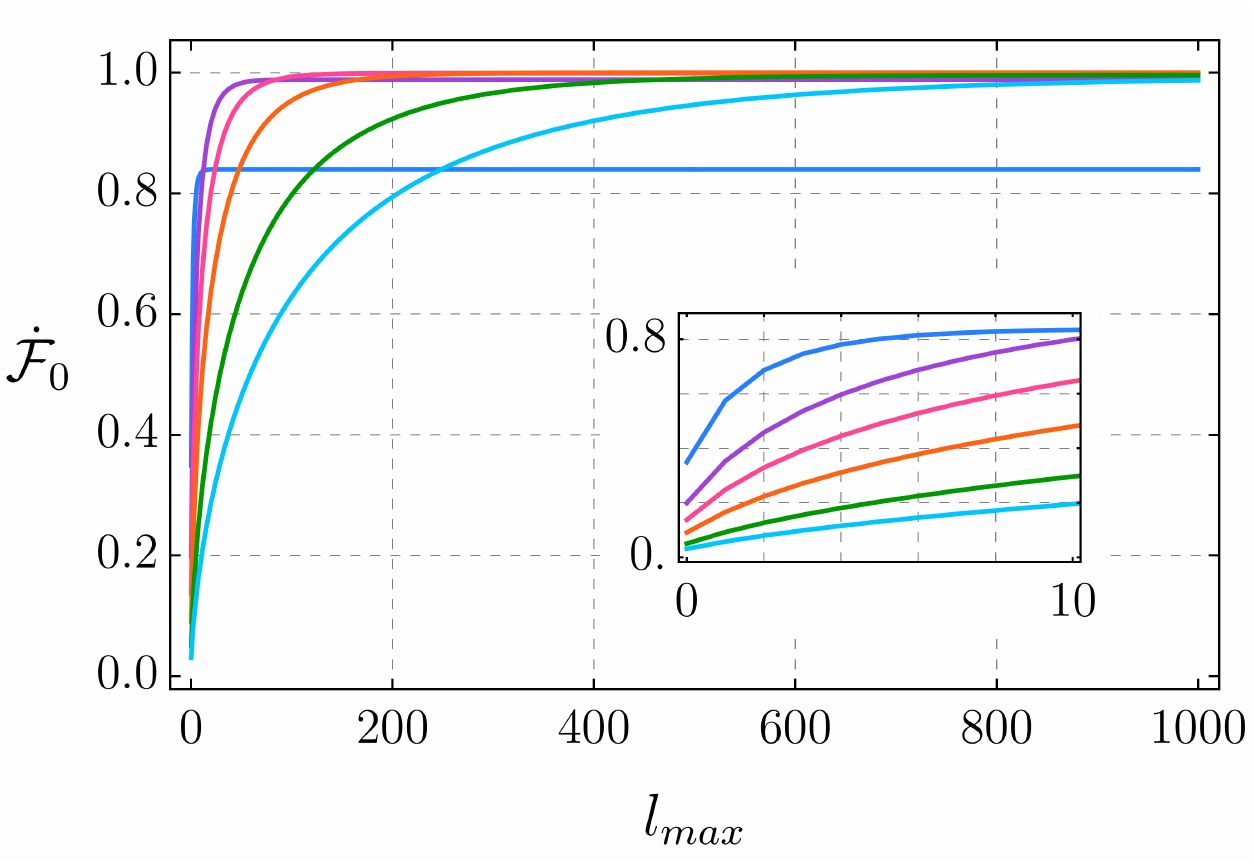}
				\caption{Transition rate for the ground state summed over the natural numbers up to $\ell_{max}$ at several temperatures for $r_h=1$, $\Omega=-0.1$ and Dirichlet boundary condition, from top to bottom in the zoom plot $T_H=(1,5,10,20,50,100)$. It indicates that the transition rate converges to a monotonically increasing function of temperature}
				\label{fig:unddfdfderst1}
				\end{figure}
\begin{figure}[H]
				\centering
						\includegraphics[width=.45\textwidth]{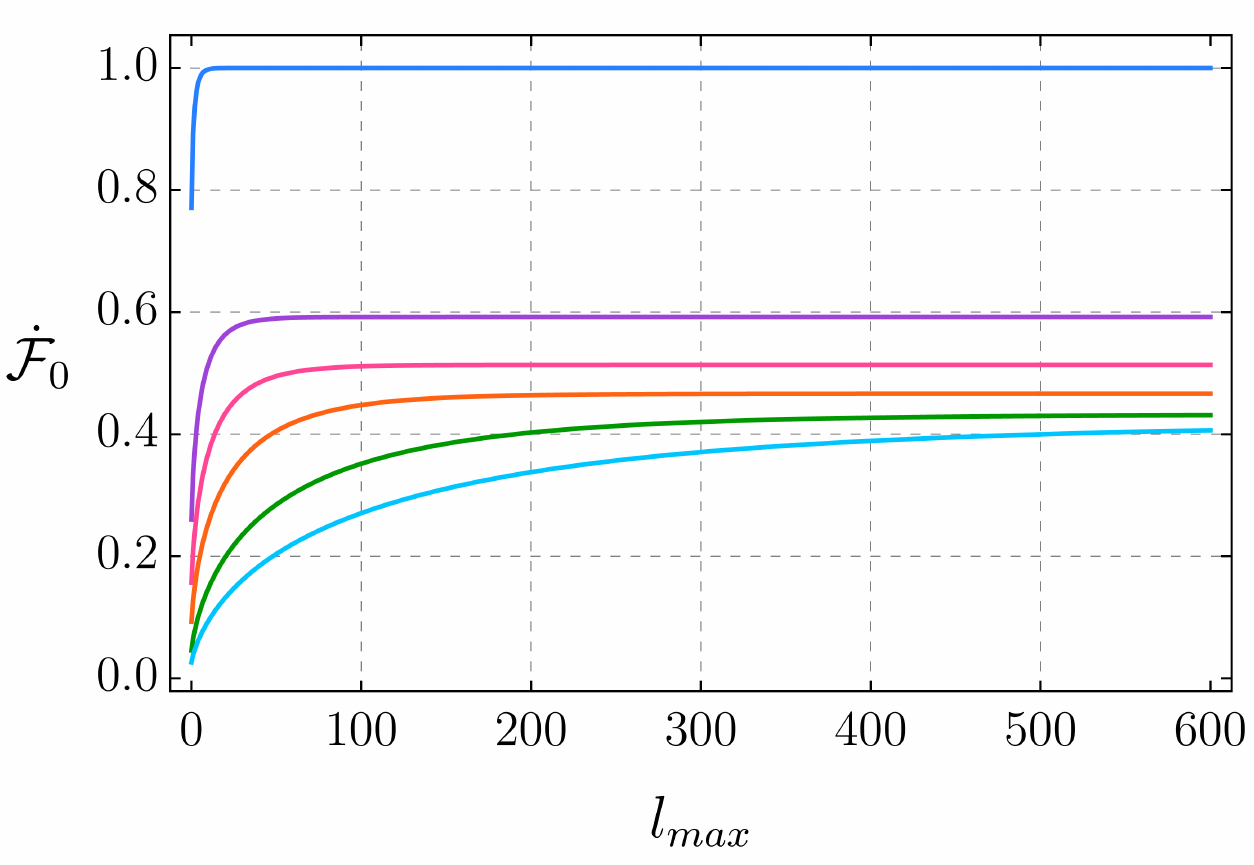}
				\caption{Transition rate for the ground state summed over the natural numbers up to $\ell_{max}$ at several temperatures for $r_h=1$, $\Omega=-0.1$ and Neumann boundary condition, from top to bottom $T_H=(1,5,10,20,50,100)$. The curves are manifestly separated, and curves of smaller temperatures remain higher than curves of higher temperatures.}
				\label{fig:unddfdfderst2}
				\end{figure}

			Our analysis strongly suggests that both the anti-Unruh effect and the anti-Hawking effect are still manifest after performing the $\ell$-sum for an horizon lying at $r_h=1$ and Neumann boundary condition. For Dirichlet boundary condition, on the contrary, we observe that the effects could be cancelled by performing the $\ell$-sum. To illustrate this, we have included Figures \ref{fig:unddfdfderst1} and \ref{fig:unddfdfderst2} showing the transition rate seen as a function of the number of $\ell$-terms which are summed ($\ell_{max}$).

			Note that the figures above show only the discrete sum over the natural numbers. However, including the negative integers $\ell$-terms for the BTZ case or numerically integrating the transition rate over real numbers for the Rindler-AdS$_3$ case, does not alter the conclusion.


		\paragraph{The scenario $r_h=0.1$}

\begin{figure}[H]
				\centering
						\includegraphics[width=.45\textwidth]{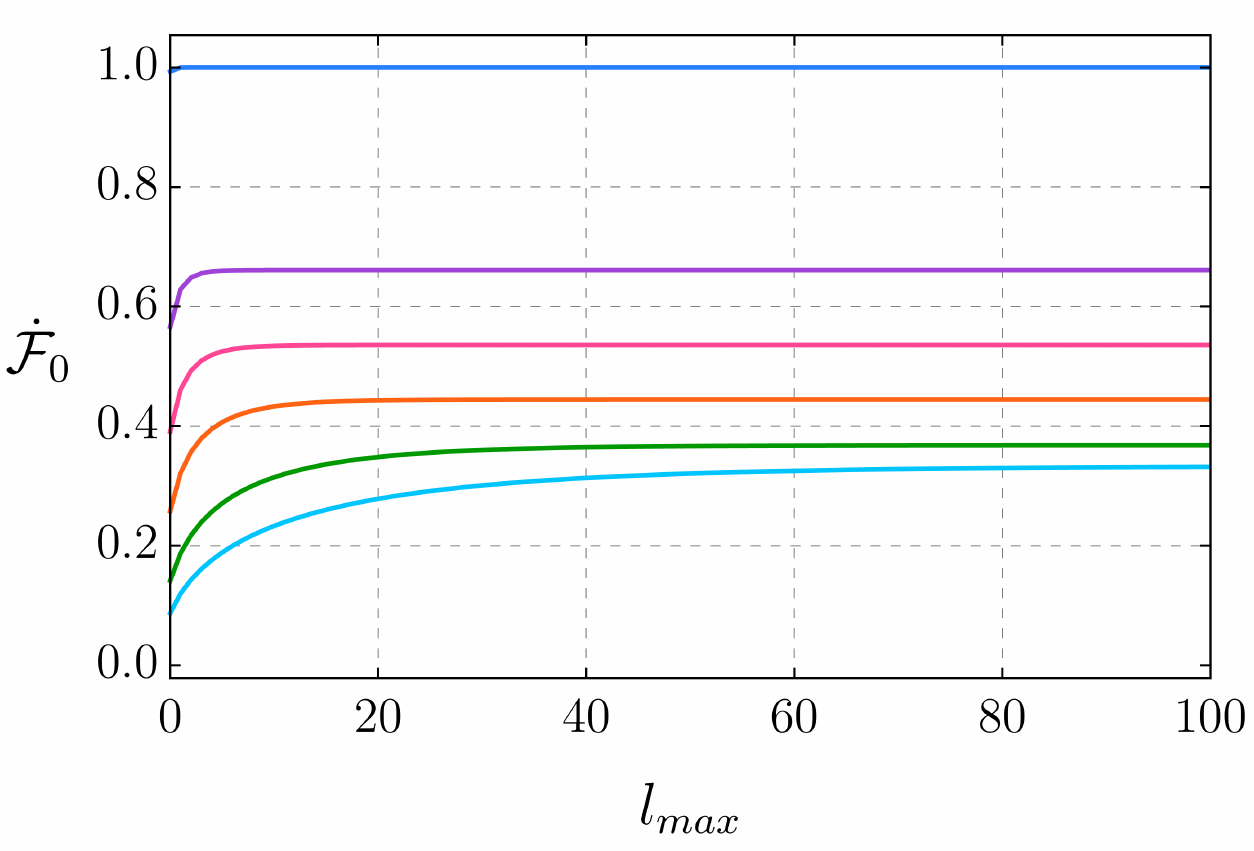}
				\caption{Transition rate for the ground state summed over the natural numbers up to $\ell_{max}$ at several temperatures for $r_h=0.1$, $\Omega=-0.1$ and Dirichlet boundary condition, from top to bottom $T_H=(1,5,10,20,50,100)$. }
				\label{fig:uuBTZ}
				\end{figure}
\begin{figure}[H]
				\centering
						\includegraphics[width=0.45\textwidth]{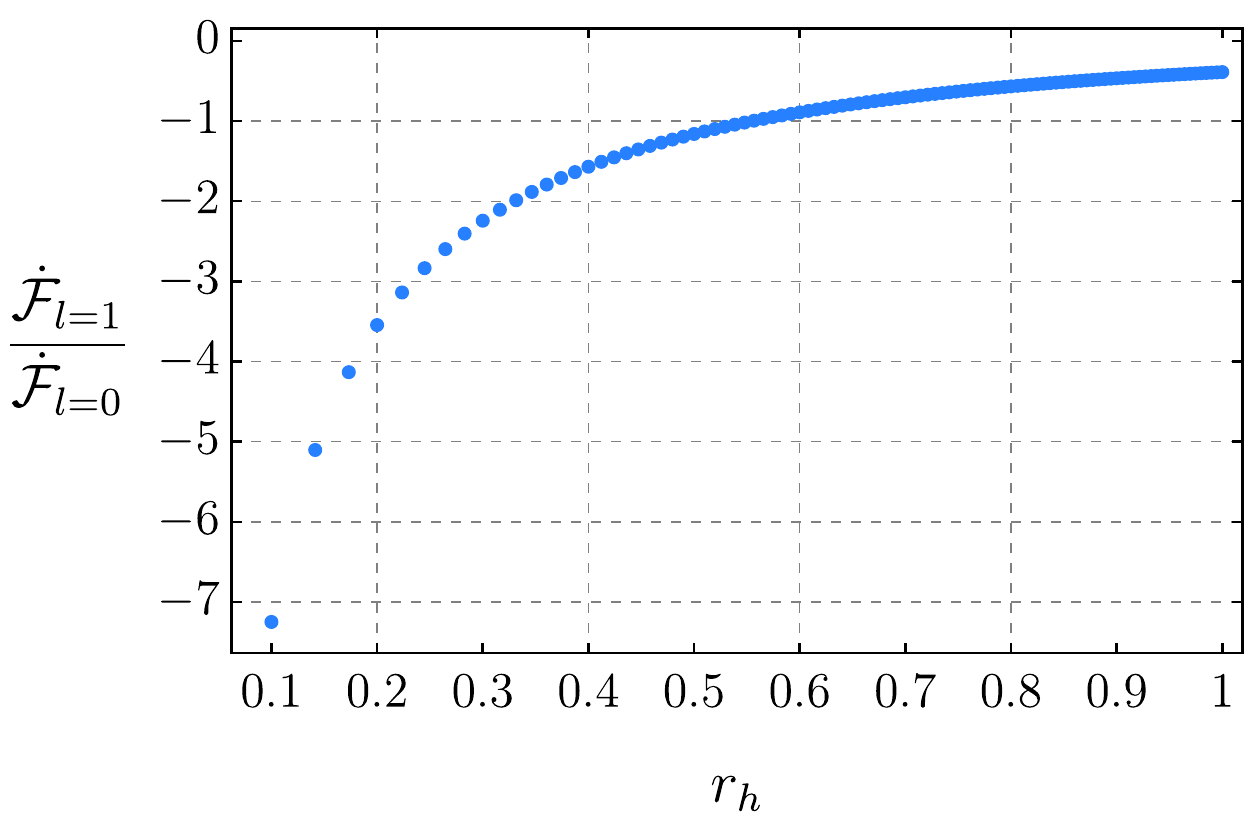}
				\caption{Logarithmic plot of the $\ell=1$ term of the transition rate as a function of the BTZ mass for the ground state with $\Omega=-0.1$, $\gamma=0$, $z_{\mini{D}}=0.5$, normalized with respect to the $\ell=0$ term. }
				\label{fig:BTZlsoverl0}
				\end{figure}

		 	For sufficiently small $r_h$, such as $r_h=0.1$, the $\ell=0$ term becomes by far the dominant one. The numerical analysis firmly indicates that for all boundary conditions the anti-Hawking effect is still manifest on a static BTZ black hole. We obtain plots analogous to Figure \ref{fig:unddfdfderst2} for both boundary conditions, such as Figure \ref{fig:uuBTZ}, which also shows a fast convergence with respect to the $\ell$-sum.
			Figure \ref{fig:BTZlsoverl0} shows the logarithmic behaviour of the transition rate for the $\ell=1$ term normalized with respect to the $\ell=0$ term for some values of $r_h$. Since its normalized, the plot holds for both the ground state and the KMS state. The aim is to illustrate that, on a BTZ spacetime, for smaller values of $r_h$, the term $\ell=0$ dominates .


\section{Conclusions}
\label{sec:conclusions}


	Our main results are the following. We confirm and we extend to all boundary conditions of Robin type the results existing in literature concerning the anti-Unruh and the anti-Hawking effect for the ground state of a real, massless, conformally coupled scalar field either on Rindler-AdS$_3$ or on a BTZ spacetime. In addition, we show that neither the anti-Unruh nor the anti-Hawking effects are manifest for the KMS states considered.

	The results we obtained concerning the ground state on Rindler-AdS$_3$ and on a static BTZ black hole are compatible with the results of \cite{Henderson:2019uqo}, where it was shown that for the ground state of the massless conformally coupled scalar field on the three-di\-men\-sion\-al AdS spacetime, an Unruh-DeWitt detector following Rindler trajectories, seeing a horizon at $r_h=1$, would experience the anti-Unruh effect for Neumann boundary condition, but not for Dirichlet or transparent boundary conditions. On a static BTZ spacetime, with a sufficiently small black hole mass, the anti-Hawking effect would be manifest for all three of these boundary conditions. Nevertheless, the framework invoked here considers more general boundary conditions and it allows to extend the analysis also to the case of massive arbitrarily coupled real scalar fields.


\section{Acknowledgments}


We are grateful to Jorma Louko and to Benito A. J. Aubry for fruitful discussions and comments on the manuscript. The work of L.S.C is supported by a PhD scholarship of the University of Pavia, which is gratefully acknowledged. Both authors also acknowledge the INFN for travel support, which played a key role in the realization of this project.

\end{multicols}


\end{document}